\documentclass[12pt]{article}
\usepackage{graphicx}

      \usepackage[cp1251]{inputenc} %% Windows
      \usepackage[russian]{babel}   %%

\begin{document}
 \noindent {\footnotesize\it Astronomy Letters, 2013, Vol. 39, No. 3, pp. 185--191}
\newcommand{\dif}{\textrm{d}}

 \noindent
 \begin{tabular}{llllllllllllllllllllllllllllllllllllllllllllll}
 & & & & & & & & & & & & & & & & & & & & & & & & & & & & & & & & & & & & & \\\hline\hline
 \end{tabular}

 \vskip 1.0cm
 \centerline {\large\bf Corrections for the Lutz–Kelker Bias for Galactic Masers}
 \bigskip
 \centerline {A.S. Stepanishchev$^1$ and V.V. Bobylev$^{1,2}$}
 \medskip
{\small\it
 $^1$~Pulkovo Astronomical Observatory, Russian Academy of Sciences

 $^2$~Sobolev Astronomical Institute, St. Petersburg State University, Russia

 }

 \bigskip
{\bf Abstract}---Based on published data, we have collected
information about Galactic maser sources with measured distances.
In particular, 44 Galactic maser sources located in star-forming
regions have trigonometric parallaxes, proper motions, and radial
velocities. In addition, ten more radio sources with incomplete
information are known, but their parallaxes have been measured
with a high accuracy. For all 54 sources, we have calculated the
corrections for the well-known Lutz-Kelker bias. Based on a sample
of 44 sources, we have refined the parameters of the Galactic
rotation curve. Thus, at $R_0=8$~kpc, the peculiar velocity
components for the Sun are
 $(U_\odot,V_\odot,W_\odot)=(7.5,17.6,8.4)\pm(1.2,1.2,1.2)$~km s$^{-1}$
and the angular velocity components are
 $\omega_0 = -28.7\pm0.5$~km s$^{-1}$ kpc$^{-1}$,
 $\omega'_0 = +4.17\pm0.10$~km s$^{-1}$ kpc$^{-2}$, and
 $\omega''_0 = -0.87\pm0.06$~km s$^{-1}$ kpc$^{-3}$.
 The corresponding Oort constants are
 $A= 16.7\pm0.6$~km s$^{-1}$ kpc$^{-1}$ and
 $B=-12.0\pm1.0$~km s$^{-1}$ kpc$^{-1}$; the circular
rotation velocity of the solar neighborhood around the Galactic
center is $V_0=230\pm16$~km s$^{-1}$. We have found that the
corrections for the Lutz-Kelker bias affect the determination of
the angular velocity $\omega_0$ most strongly; their effect on the
remaining parameters is statistically insignificant. Within the
model of a two-armed spiral pattern, we have determined the
pattern pitch angle
 $i=-6.^\circ5$ and the phase of the Sun in the spiral wave $\chi_0 = 150^\circ$.

\section{INTRODUCTION}

The observed velocities of various young Galactic objects (HI
clouds, OB stars, young open star clusters) are used to study the
kinematics of the Galaxy and the spiral structure of the Galactic
disk (Bobylev et al. 2008).

For this task, Galactic maser sources with trigonometric
parallaxes measured by VLBI are extremely important (Reid et al.
2009a; McMillan and Binney 2010; Bobylev and Bajkova 2010;
Stepanishchev and Bobylev 2010). We have in mind the maser sources
associated with the youngest Galactic stellar objects (either
protostellar objects of various masses, or very massive
supergiants, or T Tauri stars). The number of measured parallaxes
for such masers already exceeds 50 and with a complete data set
(with proper motions and radial velocities) is 44.

Studying their distribution and kinematics with allowance made for
the distance errors is of great importance. First of all, it is
necessary to take into account the Lutz–Kelker (1973) bias. The
corrections for this bias usually play an important role in
determining the trigonometric parallaxes by optical methods, when
the reference objects are distant stars rather than extragalactic
sources. In this case, a procedure for absolutizing the measured
relative parallaxes is needed. For this purpose, the corrections
calculated from the Lutz–Kelker bias are usually applied.

In the case of Galactic maser sources, the situation is different.
Their trigonometric parallaxes are determined by VLBI with
referencing directly to distant quasars (two or three quasars are
commonly used). Therefore, such parallaxes are immediately
obtained as absolute ones.

At the same time, the VLBI parallaxes of masers were measured with
a certain random error that is, on average, very small, about 5\%.
However, there are exceptions and, in several cases, the error
exceeds 20\%. As is well known (Lutz and Kelker 1973), the
knowledge of the true parallax depends strongly on the relative
error of the measured parallax. It can be surmised that the
parameters of the Galactic rotation curve and the parameters of
the spiral density wave determined from maser sources can be
distorted because of the uncertainties in their distances.

Great experience in taking into account the Lutz-Kelker bias has
been gained. We will point out only some of the papers.
Hanson~(1979) applied the method in practice to calibrate the
luminosity based on trigonometric parallaxes. In the literature,
the method is occasionally called the Lutz-Kelker-Hanson method.
Ma\'iz-Apell\'aniz~(2001) determined such corrections for O-B5
stars with trigonometric parallaxes from Hipparcos (1997), with
the Sun’s height above the Galactic plane having been taken into
account. Verbiest et al. (2010) proposed a modification of the
method that took into account the luminosity function as applied
to 57 Galactic pulsars with measured trigonometric parallaxes.
Note that the Lutz–Kelker bias is a statistical effect. A very
accurate model of the expected distribution of stars in the Galaxy
is required to properly take the Lutz-Kelker bias into account.

The goal of this paper is to calculate the corrections for the
Lutz–Kelker bias for masers with measured trigonometric parallaxes
and to study the effect of these corrections on the parameters of
the Galactic rotation curve and the parameters of the spiral
density wave being determined.

\section{METHOD}
\subsection{Uniform Distribution in Space}

In its classical form, the Lutz–Kelker bias refers to a uniform
distribution of stars in an infinite space (Lutz and Kelker 1973).
We assume that the errors in the distribution of the measured
parallax relative to its true value are distributed normally:
\begin{equation}
p(\varpi_0|\varpi)=\frac{1}{\sqrt{2\pi}\sigma}
\exp\left(-\frac{(\varpi_0-\varpi)^2}{2\sigma^2}\right).
\end{equation}
Here, $\pi=3.14\ldots$, $\sigma$ у is the measurement error of the
parallax, $\varpi_0$ and $\varpi$ are the measured and true
parallaxes, respectively. We are interested in the distribution of
true parallaxes $\varpi$ if a sufficient number of stars with a
certain measured value of $\varpi_0$ are available. If the stars
are uniformly distributed in space, then their number in an
interval $(r,r+\dif r)$ is
\begin{equation}
N(r)\dif r=4\pi r^2\dif r,
\end{equation}
or
\begin{equation}
N(\varpi)\dif\varpi=\frac{4\pi\dif\varpi}{\varpi^4}.
\end{equation}
According to the Bayes theorem,
\begin{equation}
p(\varpi|\varpi_0)=\frac{p(\varpi_0|\varpi)p(\varpi)}{p(\varpi_0)},
\end{equation}
where $p(\varpi_0)$ may be set equal to a constant. We find that
\begin{equation}\label{eq_main}
p(\varpi|\varpi_0)=\frac{\kappa}{\varpi^4}
\exp\left(-\frac{(\varpi_0-\varpi)^2}{2\sigma^2}\right).
\end{equation}
The constant $\kappa$ is chosen in such a way that the
normalization condition is fulfilled:
\begin{equation}\label{eq_norm}
\int\limits_{\varpi_{min}}^\infty p(\varpi|\varpi_0)\dif\varpi=1.
\end{equation}
The lower limit $\varpi_{min}$ implies the maximum distance at
which a given star can be observed in principle from the
considerations of the stellar system’s size and the telescope’s
limiting magnitude. In the case of integration from zero, the
integral diverges. This reflects an infinite number of stars in an
infinite space for the chosen uniform distribution.

The parallax corrected for the Lutz–Kelker bias is the expectation
of distribution (\ref{eq_main}):
\begin{equation}\label{eq_corr}
\varpi_{corr}=\int\limits_{\varpi_{min}}^\infty \varpi
p(\varpi|\varpi_0)\dif\varpi.
\end{equation}

\subsection{An Exponential Disk with an Observer at the Periphery}

Our objective is to study the objects distributed in the Galactic
disk. We assume an exponential density distribution along the
Galactic radius, with the distribution along the vertical $Z$ axis
following the law of the hyperbolic secant squared:
\begin{equation}
\rho(R)=\rho_0\exp\left(-\frac{R}{h}\right)\ch^{-2}\left(\frac{Z}{Z_0}\right),\quad
h=3\textrm{ kpc}, \quad Z_0=0.25\textrm{ kpc},
\end{equation}
where $R$ is the Galactocentric distance, $h$ and $Z_0$ are the
radial and vertical scale lengths of the disk, respectively; the
constant $\rho_0$, the density at the center, is a scale factor
and its value is of no importance for our problem. The Sun is at
$R_0=8$~kpc from the Galactic center. When observing from the Sun
along the Galactic longitude $l$, the density distribution takes
the form
\begin{equation}
 \rho(r,l)=\rho_0\exp\left(-\frac{\sqrt{(r\cos l-R_0)^2+r^2\sin^2 l}}{h}\right).
\end{equation}
The parallax corrected for the Lutz–Kelker bias is calculated from
(\ref{eq_corr}) with the distribution of true parallaxes at given
observables:
\begin{equation}
p(\varpi|\varpi_0)\propto\frac{\rho(r,l)}{\varpi^4}\exp
\left(-\frac{1}{2}\frac{(\varpi-\varpi_0)^2}{\sigma^2}\right).
\end{equation}

%%%%%%%%%%%%%%%%%%%%%%%%%%%%%%%%%%%%%
      \begin{table}
      \begin{center}
 \caption{Data on the sources that were not included in the ``kinematic'' sample}
      {\small
      \begin{tabular}{|l|r|r|r|r|r|r|r|r|r|r|r|}      \hline
      Source & $\alpha$ & $\delta$ & $\varpi(\sigma_\varpi)$ &
      $\mu^*_\alpha (\sigma_{\mu_\alpha})$ & $\mu_\delta(\sigma_{\mu_\delta})$ &
      $V_r(\sigma_{V_r})$ & Ref \\\hline

 G~23.66$-$0.13    & $278.7149$ & $ -8.3059$ & $ .313 (.039)$ & $-1.32 (.02)$ & $-2.96 (.03)$ & $ 83  (3)$ &(a) \\\hline % Bartkev,2008 \\\hline
 G~9.62$+$0.20     & $271.5611$ & $-20.5255$ & $ .194 (.023)$ & $ -.58 (.05)$ & $-2.49 (.27)$ & $0.1  (2)$ &(b) \\\hline % Sanna,  2009 \\\hline
 Sgr B2N           & $266.8330$ & $-28.3720$ & $ .128 (.015)$ & $ -.32 (.05)$ & $-4.69 (.11)$ & $ 64  (5)$ &(c) \\\hline % Reid,   2009 \\\hline
 Sgr B2M           & $266.8340$ & $-28.3845$ & $ .130 (.012)$ & $-1.23 (.04)$ & $-3.84 (.11)$ & $ 61  (5)$ &(c) \\\hline % Reid,   2009 \\\hline
 IRAS~20126$+$41   & $303.6084$ & $ 41.2257$ & $ .61  (.02 )$ & $-2.0  (.1 )$ & $ 1.0  (.1 )$ & $-3.5 (4)$ &(d) \\\hline % Moscadel2010 \\\hline
 G~48.61$+$0.02    & $290.1299$ & $ 13.9237$ & $ .199 (.007)$ & $-2.76 (.04)$ & $-5.28 (.11)$ & $ 19  (1)$ &(e) \\\hline % Nagayama2011 \\\hline
 MSXDC G034.4$+$0  & $283.3292$ & $  1.4022$ & $ .643 (.049)$ & $ -.25 (.18)$ & $  .00 (.18)$ & $ 57  (5)$ &(f) \\\hline % VERA    2011 \\\hline
 EC 95             & $277.4912$ & $  1.2128$ & $2.41  (.02 )$ & $  .70 (.02)$ & $-3.64 (.10)$ &        --- &(g) \\\hline % Dzib,et 2010 \\\hline
 IRAS~05137$+$3919 & $ 79.3073$ & $ 39.3722$ & $ .086 (.027)$ & $  .30 (.10)$ & $ -.89 (.27)$ & $-26  (3)$ &(h) \\\hline %   VERA    2011  \\\hline
 G~27.36$-$0.16    & $280.4627$ & $ -5.0287$ & $ .125 (.042)$ & $-1.81 (.08)$ & $-4.11 (.26)$ & $92.2 (5)$ &(i) \\\hline %   Mosc NEW2011  \\\hline

       \end{tabular}}
      \end{center}
{\small Note. $\alpha$ and $\delta$ are given in degrees, the
parallax $\varpi$ is in mas, $\mu^*_\alpha=\mu_\alpha\cos\delta$
and $\mu_\delta$ are in mas yr$^{-1}$, the radial velocity
$V_r=V_r(LSR)$ is in km s$^{-1}$; the letters mark the references
to papers: (a)~Bartkiewicz et al. (2008); (b)~Sanna et al. (2009);
(c)~Reid et al. (2009); (d)~Moscadelli et al. (2011); (e)~Nagayama
et al. (2011b); (f)~Kurayama et al. (2011); (g)~Dzib et al.
(2010); (h)~Honma et al. (2011); (i)~Xu et al. (2011).      }
      \end{table}
%%%%%%%%%%%%%%%%%%%%%%%%%%%%%%%%%%%

\section{DATA}

The trigonometric parallaxes and proper motions of Galactic maser
sources are determined by several research groups using long-term
VLBI observations. These include the Japanese VERA (VLBI
Exploration of Radio Astrometry) project on the observation of
Galactic H2O maser sources at 22 GHz and SiO masers (these are
very few among the young objects) at 43 GHz. Note that the higher
the frequency, the higher the resolution and the more accurate the
observations. Methanol (CH3OH) masers are observed at 12 GHz using
the American NRAO VLBA and at 6.7 GHz using the European VLBI
Network. At present, Galactic masers are observed by these groups
of researchers within the unified BeSSeL project (Brunthaler et
al. 2011) aimed at studying the Galactic structure. VLBI
observations of a number of young radio stars in continuum at 8.4
GHz (Dzib et al. 2010) are also conducted to study the structure
and kinematics of the solar neighborhood (the Gould Belt,
molecular clouds).

Bobylev and Bajkova (2010) and Stepanishchev and Bobylev (2011)
analyzed a sample of 28 masers with measured trigonometric
parallaxes drawn from published data. By now, the amount of such
data has increased considerably---about 20 new measurements in
various star-forming regions have been published. First of all, we
use the input data on 44 masers with measured trigonometric
parallaxes, proper motions, and radial velocities from Bajkova and
Bobylev (2012), where corresponding references and explanations
are given. We arbitrarily call this sample ``kinematic'', because
on its basis we determine the Galactic rotation parameters.

Table 1 gives the parameters of the sources most of which have
small relative errors in the parallaxes, but, for various reasons,
they were not included in the ``kinematic'' sample. Two sources
are almost at the Galactic center, Sgr B2N and Sgr B2M (Read et
al. 2009a). Other sources have large (>40 km s$^{-1}$) residual
velocities: these include G23.66.0.13 (Bartkiewicz et al. 2008),
MSXDC G034.43+00.24 (Kurayama et al. 2011), G48.61+0.02 (Nagayama
et al. 2011), and IRAS 20126+4104 (Moscadelli et al. 2011). Since
the star-forming region G9.62+0.20 (Sanna et al. 2009) is closely
connected with the 3-kpc spiral arm, it has a large peculiar
Galactocentric radial velocity, $|V_R|\approx50$ km s$^{-1}$. In
addition, we included the radio star EC~95 (Dzib et al. 2010) in
our sample, for which there are no radial velocity measurements as
yet. Its VLBI observations were carried out in continuum at
8.42~GHz. It is hypothesized that this is a binary system
consisting of a Herbig-Haro protostar with a mass of about
$4-5~M_\odot$ and a T Tauri low-mass companion.

The stars from Table 1, especially those farthest from the Sun,
are of interest in studying the effect of the Lutz–Kelker bias on
the determination of the spiral-structure parameters.

 \begin{figure}[t] {\begin{center}
 \includegraphics[width= 100mm]{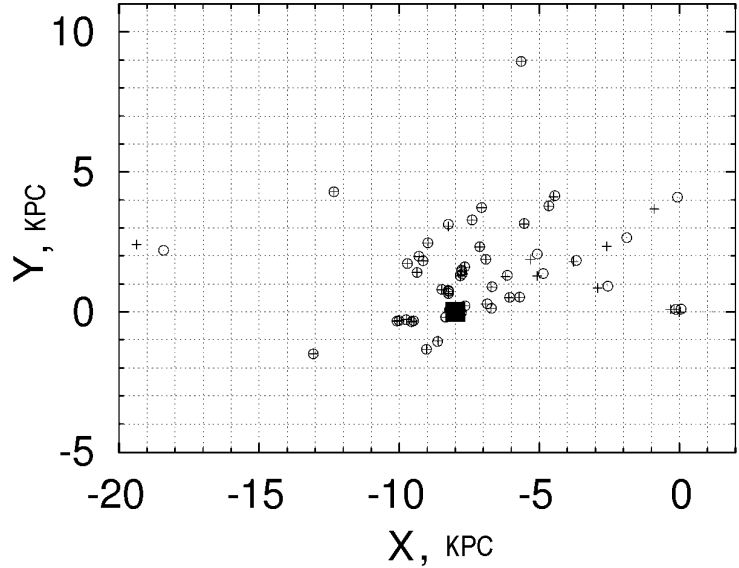}
 \caption{Projections of masers onto the Galactic plane
$(XY)$ from the observed data (crosses) and corrected for the
Lutz-Kelker bias (circles); the large filled square marks the
position of the Sun; the Galactic Center is at the coordinate
origin.}
 \label{f-1}
 \end{center} }
 \end{figure}

 \begin{figure}[t] {\begin{center}
 \includegraphics[width= 100mm]{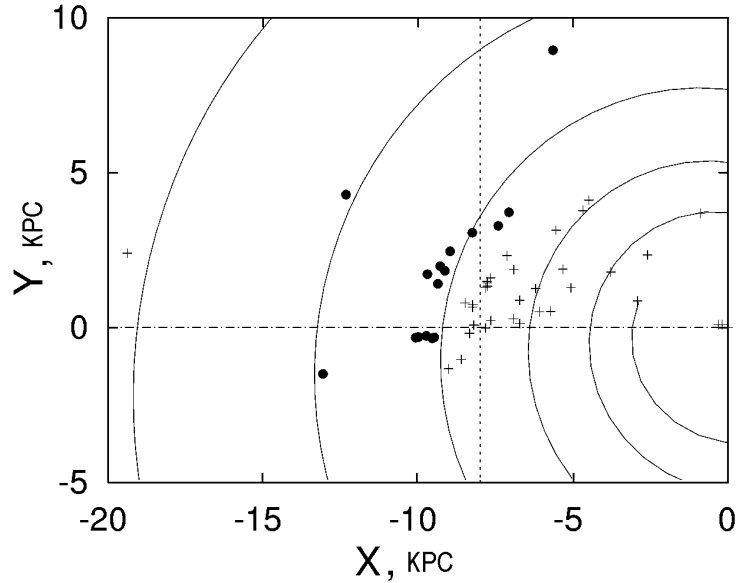}
 \caption{Two-armed spiral pattern constructed from
masers in the Perseus and Outer Arms (thick circles); the
remaining objects from our list are indicated by the crosses.}
 \label{f-1}
 \end{center} }
 \end{figure}

\section{RESULTS AND DISCUSSION}

The observed ($\varpi_{obs}$) and Lutz-Kelker-biascorrected
($\varpi_{corr}$) parallaxes for all 54 maser sources are given in
Table 2; their distribution in projection onto the Galactic $XY$
plane is shown in Figs. 1 and 2. As can be seen from Table 2, for
most of the maser sources, the calculated Lutz-Kelker bias
corrections are negative (the heliocentric distances to them
should be increased). The correction is largest for IRAS
16293$-$2422, but this source is one of the closest to the Sun.
Therefore, the correction found has no critical effect on the
determination of the (distance-dependent) parameters under
consideration. As can be seen from Fig. 1, a significant change in
the distance is achieved for several sources near the Galactic
center. The source farthest from the Sun, IRAS 05137+39, is
greatest interest. The relative error in the parallax for it is
$\approx$30\%; since the source is close to the direction of the
Galactic anticenter, the correction for the Lutz-Kelker bias is
positive (its heliocentric distance should be reduced).

Figure 1 clearly illustrates the essence of the method. More
specifically, the existence of small measurement errors in the
parallaxes gives grounds to expect that the most probable distance
to a star should be closer to the region with the highest density
of stars (in our case, this is the Galactic center region).

\subsection{Spiral Pattern}

The parameters of the spiral pattern were determined by the method
described in Stepanishchev and Bobylev (2011). The essence of the
method consists in the following. We proceed from the well-known
equation (Lin and Shu 1964)
\begin{equation}
\chi=m(\ctg i\cdot\ln\{R/R_0\}-\theta)+\chi_0, \label{spiralequat}
\end{equation}
where $\chi$ is the phase of the object in the spiral density
wave, $\chi_0$ is the phase of the Sun in the spiral wave, $R$ is
the distance from the object to the Galactic rotation axis,
$\theta$ is the Galactocentric longitude of the object, and $m$ is
the number of spiral arms (here, we take $m=2$). Having chosen two
or three groups of objects located along the arms, we estimate the
most probable phase of the Sun and the pitch angle of the spiral
pattern defined by Eq. (11). For this purpose, we search for the
minimum of the functional
\begin{equation}
\zeta^2(i,\chi_0)=\sum (p_j\chi_j)^2,
\end{equation}
where $\chi_j$ is the phase of the $j$th object reduced to the
range $(-\pi/2,\pi/2)$ and $p_j$ is the weight that is chosen
based on the number of objects in a given arm in such a way that
all arms have the same weight. Here, the masers are assumed to be
theoretically concentrated to the spiral arm centers, i.e., the
sum of the squares of the objects’ phases in the density wave is
minimized by taking into account the fact that the phase
$\chi=0^\circ$. corresponds to the arm center.

Owing to the increase in the amount of observational data, these
parameters can be calculated with a higher accuracy. Thus, three
masers are already known in the Outer Arm against two in the
previous paper, while eleven masers are known in the Perseus Arm
against seven. In Fig. 2, the objects involved in choosing the
model parameters are indicated by the filled circles and the
remaining ones are indicated by the crosses. The pitch angle and
the Sun's phase were found to be, respectively, $i=-6^\circ.5$ and
$\chi_0=150^\circ$.

Since the Lutz-Kelker bias corrections for most of the masers
belonging to the spiral arms (from which the parameters of the
spiral pattern were determined) are insignificant, they were
disregarded in the case under consideration. Note once again the
source IRAS 05137+39, which is behind the Outer Arm (at
$\approx$20~kpc from the Galactic center). It was not used in
choosing the model parameters; nevertheless, it falls nicely on
the logarithmic spiral fitted into the Perseus and Outer Arms
(Fig. 1 and Fig. 2). In their kinematic analysis of a sample of
Galactic masers, Bajkova and Bobylev (2012) found a slightly
smaller pitch angle of the spiral wave,
$i=-5^{+0.2^\circ}_{-0.9^\circ}$ (for $m=2$), and a phase of the
Sun in the spiral wave fairly close to our result,
$\chi_\odot=-147^{+3^\circ}_{-17^\circ}$  (measured from the
Carina-Sagittarius Arm). At the same time, the spiral pitch angle
we derived differs significantly from the result of analyzing the
samples of masers in the Perseus and Outer Arms obtained by Sakai
et al. (2012) based on a different method (with a separate
analysis of each arm). These authors found
$i=-17^\circ.8\pm1^\circ.7$ for the Perseus Arm and
$i=-11^\circ.6$ for the Outer Arm. This discrepancy in determining
the pitch angle can be explained by the fact that the results of
Sakai et al. (2012) more closely correspond to the four-armed
model of the spiral pattern (in our case, we will then have
$i=-13^\circ$). On the whole, we can conclude that considerably
larger statistics is required for a reliable determination of the
spiral pattern parameters.

\subsection{Galactic Rotation Parameters}

First, we redetermined the parameters of the Galactic rotation
curve using data on 44 masers. Our technique of its determination
based on a Taylor expansion of the angular velocity in terms of
the Galactocentric distance was described in detail in Bobylev and
Bajkova (2010) and Bobylev and Stepanishchev (2011). At a fixed
Galactocentric distance of the Sun, $R_0=8$~kpc (with an
uncertainty of 0.5~kpc), the peculiar velocity components of the
Sun
 $(U_\odot,V_\odot,W_\odot)=(7.5,17.6,8.4)\pm(1.2,1.2,1.2)$~km s$^{-1}$
 and the following parameters of the Galactic rotation
curve were obtained:
\begin{equation}
 \begin{array}{rllcc}
 \omega_0   &=&-28.7 \pm0.5 ~\hbox{~km s$^{-1}$ kpc$^{-1}$},\\
 \omega'_0  &=& +4.17\pm0.10~\hbox{~km s$^{-1}$ kpc$^{-2}$},\\
 \omega''_0 &=& -0.87\pm0.06~\hbox{~km s$^{-1}$ kpc$^{-3}$},
 \label{44}
 \end{array}
 \end{equation}
where  $\omega_0$ is the angular velocity of Galactic rotation at
$R=R_0,$ $\omega'_0$ and $\omega''_0$ are its corresponding
derivatives. The Oort constants are $A=16.7\pm0.6$~km s$^{-1}$
kpc$^{-1}$ and $B=-12.0\pm1.0$~km s$^{-1}$ kpc$^{-1}$; the
circular rotation velocity of the solar neighborhood around the
Galactic center is $V_0 = 230\pm16$~km s$^{-1}$.

Note that Stepanishchev and Bobylev (2011) found the analogous
parameters using only 28 masers to be the following:
$(U_\odot,V_\odot,W_\odot)=(8.5,17.1,8.9)\pm(1.6,1.6,1.6)$~km
s$^{-1}$, and
 $\omega_0 =-30.4\pm0.7$~km s$^{-1}$ kpc$^{-1}$,
 $\omega'_0= 4.23\pm0.13$~km s$^{-1}$ kpc$^{-2}$,
 $\omega''_0=-1.01\pm0.06$~km s$^{-1}$ kpc$^{-3}$. It can be seen that solution (13) is more reliable,
because, as would be expected, the random errors of all the
parameters being determined decreased. Solution (13) virtually
coincides with results of analyzing this sample of masers from
Bajkova and Bobylev (2012), where a detailed comparison of these
values with the results of other authors can be found.

After applying the Lutz–Kelker correction, we obtained the
following parameters of the rotation curve:
$(U_\odot,V_\odot,W_\odot)=(6.6,17.9,8.5)\pm(1.4,1.4,1.4)$~km
s$^{-1}$ and
 \begin{equation}
 \begin{array}{rllcc}
 \omega_0   &=&-27.5 \pm0.5~  \hbox{km s$^{-1}$ kpc$^{-1}$},\\
 \omega'_0  &=& +3.98\pm0.11~ \hbox{km s$^{-1}$ kpc$^{-2}$},\\
 \omega''_0 &=& -0.83\pm0.06~ \hbox{km s$^{-1}$ kpc$^{-3}$},
 \end{array}
 \end{equation}
Hence we see that after applying the correction, the angular
velocity $\omega_0$ changed by about $2.5\sigma$, while its
derivatives and the Sun’s peculiar velocity components change
within the error limits (i.e., insignificantly).

\section{CONCLUSIONS}

Based on published data, we collected information about Galactic
maser sources with well measured distances. In particular, 44
Galactic maser sources located in star-forming regions have a
complete set of kinematic observational data--the VLBI
trigonometric parallaxes and proper motions; the mean radial
velocities derived from CO observations are also known for them.
In addition, ten more radio sources with incomplete information
are known, but their parallaxes were measured with a sufficiently
high accuracy.

We calculated the corrections for the well-known Lutz–Kelker bias
for all 54 sources. For this purpose, we used a model with an
exponential star density distribution in the Galactic disk. For
several sources farthest from the Sun, these corrections were
shown to be appreciable. Using the calculated corrections can be
useful, for example, in studying the membership of masers in the
Galactic bar or specific spiral arms, etc.

Based on a sample of 44 sources, we refined the parameters of the
Galactic rotation curve. At R0 = 8kpc, we determined the peculiar
velocity components of the Sun
 $(U_\odot,V_\odot,W_\odot)=(7.5,17.6,8.4)\pm(1.2,1.2,1.2)$~km s$^{-1}$ and the following parameters:
 $\omega_0=-28.7\pm0.5$~km s$^{-1}$ kpc$^{-1}$,
 $\omega'_0 = +4.17\pm0.10$~km s$^{-1}$ kpc$^{-2}$ и
 $\omega''_0= -0.87\pm0.06$~km s$^{-1}$ kpc$^{-3}$.
 The corresponding Oort constants are
  $A = 16.7 \pm0.6$~km s$^{-1}$  kpc$^{-1}$ and
  $B = -12.0\pm1.0$~km s$^{-1}$  kpc$^{-1}$;
  the circular rotation
velocity of the solar neighborhood around the Galactic center is
 $V_0 = 230\pm16$~km s$^{-1}$. We found that when using our sample, the
corrections for the Lutz-Kelker bias affect the determination of
the angular velocity of Galactic rotation $\omega_0$ (at the
$2.5\sigma$ level), while these corrections have no statistically
significant effect on the other model parameters being determined.
This is primarily due to the high accuracy of the measured
trigonometric parallaxes for the masers used. Within the model of
a two-armed spiral pattern, we determined the pattern pitch angle
$i=-6^\circ.5$ and the phase of the Sun in the spiral wave
$\chi_0=150^\circ$.

\subsection*{ACKNOWLEDGMENTS}

This work was supported in part by the ``Origin, Structure, and
Evolution of Objects of the Universe'' Program of the Presidium of
the Russian Academy of Sciences and the Program for State Support
of Leading Scientific Schools of the Russian Federation (project
no. NSh-3645.2010.2 ``Multiwavelength Astrophysical Research'').

\subsection*{REFERENCES}

{\small
 ~~~~ 1. A.T. Bajkova and V.V. Bobylev, Astron. Lett. {\bf 38}, (2012).

 2. A. Bartkiewicz, A. Brunthaler, M. Szymczak, {\it et al.}, Astron. Astrophys. {\bf 490}, 787 (2008).

 3. V.V. Bobylev, A.T. Bajkova, and A.S. Stepanishchev, Astron. Lett. {\bf 34}, 515 (2008).

 4. V.V. Bobylev and A.T. Bajkova, Mon. Not. R. Astron. Soc. {\bf 408}, 1788 (2010).

 5. A. Brunthaler, M. Reid, K. M. Menten, {\it et al.}, Astron. Nachr. {\bf 332}, 461 (2011).

 6. S. Dzib, L. Loinard, A. J. Mioduszewski, {\it et al.}, Astrophys. J. {\bf 718}, 610 (2010).

 7. R.B. Hanson, Mon. Not. R. Astron. Soc. {\bf 186}, 875 (1979).

 8. HIPPARCOS and Tycho Catalogues, ESA SP--1200 (1997).

 9. T. Honma, T. Hirota, Y. Kan-ya, {\it et al.}, Publ. Astron. Soc. Jpn. {\bf 63}, 17 (2011).

 10. T. Kurayama, A. Nakagawa, S. Sawada-Satoh, {\it et al.}, Publ. Astron. Soc. Jpn. {\bf 63}, 513 (2011).

 11. C.C. Lin and F.H. Shu, Astroph. J. {\bf 140}, 646 (1964).

 12. T.E. Lutz and D.H. Kelker, Publ. Astron. Soc. Pacif. {\bf 85}, 573 (1973).

 13. J. Ma\'{i}z-Apell\'aniz, Astron. J. {\bf 121}, 2737 (2001).

 14. P.J. McMillan and J.J. Binney, Mon. Not. R. Astron. Soc. {\bf 402}, 934 (2010).

 15. L. Moscadelli, R. Cesaroni,M. J. Rioja, {\it et al.}, Astron. Astrophys. {\bf 526}, 66 (2011).

 16. T. Nagayama, T. Omodaka, T. Handa, {\it et al.}, Publ. Astron. Soc. Jpn. {\bf 63}, 719 (2011b).

 17. M.J. Reid, K.M.Menten, X.W. Zheng, {\it et al.}, Astrophys. J. {\bf 700}, 137 (2009a).

 18. M. Reid, K.M. Menten, X.W. Zheng, {\it et al.}, Astrophys. J. {\bf 705}, 1548 (2009b).

 19. N. Sakai, M. Honma, H. Nakanishi, {\it et al.}, Publ. Astron. Soc. Jpn. {\bf 64}, 108 (2012).

 20. A. Sanna, M.J. Reid, L. Moscadelli, {\it et al.}, Astrophys. J. {\bf 706}, 464 (2009).

 21. A.S. Stepanishchev and V.V. Bobylev, Astron. Lett. {\bf 37}, 254 (2011).

 22. J.P.W. Verbiest, D.R. Lorimer, and M.A. McLaughlin, Mon. Not. R. Astron. Soc. {\bf 405}, 564 (2010).

 23. Y. Xu, L. Moscadelli, M.J. Reid, {\it et al.}, Astrophys. J. {\bf 733}, 25 (2011).

}

\begin{table} [p]
 \begin{center}
 \caption{Observed and Lutz–Kelker-bias-corrected parallaxes (in mas)}
  \small
  \begin{tabular}{|l|r|r|r|}\hline
             Source& $\varpi_{obs}$ & $\varpi_{corr}$&Correction\\\hline
 G 121.28$+$0.65   &  1.077 (0.039) & 1.072& -0.005\\
 IRAS 00420$+$5530 &  0.460 (0.010) & 0.460&  0.000\\
 G 123.05$-$6.31   &  0.421 (0.022) & 0.419& -0.002\\
 S Per             &  0.413 (0.017) & 0.411& -0.002\\
 G 133.94$+$1.04   &  0.512 (0.007) & 0.512&  0.000\\
 WB89$-$437        &  0.164 (0.006) & 0.164&  0.000\\
 SVS13 f1+f2       &  4.250 (0.320) & 4.158& -0.092\\
 Ori KL H2O        &  2.425 (0.035) & 2.423& -0.002\\
 Ori KL SiO        &  2.390 (0.030) & 2.389& -0.001\\
 G 189.78$+$0.34   &  0.476 (0.006) & 0.476&  0.000\\
 G 188.95$+$0.89   &  0.569 (0.034) & 0.562& -0.007\\
 G 188.78$+$1.03   &  0.496 (0.031) & 0.490& -0.006\\
 G 192.60$-$0.05   &  0.628 (0.027) & 0.624& -0.004\\
 G 196.45$-$1.68   &  0.189 (0.008) & 0.188& -0.001\\
 VY CMa            &  0.830 (0.080) & 0.803& -0.027\\
 G 232.62$+$0.99   &  0.596 (0.035) & 0.589& -0.007\\
 G 14.33$-$0.64    &  0.893 (0.101) & 0.839& -0.054\\
 G 23.43$-$0.18    &  0.170 (0.032) & 0.143& -0.027\\
 G 23.01$-$0.41    &  0.218 (0.017) & 0.211& -0.007\\
 G 35.20$-$0.74    &  0.456 (0.045) & 0.435& -0.021\\
 G 35.20$-$1.74    &  0.306 (0.045) & 0.274& -0.032\\
 IRAS 19213$+$1723 &  0.251 (0.010) & 0.249& -0.002\\
 W51 Main/South    &  0.185 (0.010) & 0.183& -0.002\\
 G 59.78$+$0.06    &  0.463 (0.020) & 0.459& -0.004\\
 G 94.58$-$1.79    &  0.326 (0.032) & 0.315& -0.011\\
 L 1204 G          &  1.309 (0.047) & 1.303& -0.006\\
 G 108.18$+$5.51   &  1.289 (0.153) & 1.217& -0.072\\
 G 109.86$+$2.10   &  1.430 (0.080) & 1.412& -0.018\\
 G 111.53$+$0.76   &  0.378 (0.017) & 0.375& -0.003\\
 IRAS~16293$-$2422 &  5.600 (1.100) & 4.425& -1.175\\
 L 1448 C          &  4.310 (0.330) & 4.214& -0.096\\
 G 5.89$-$0.39     &  0.780 (0.050) & 0.765& -0.015\\
 Onsala1           &  0.404 (0.017) & 0.401& -0.003\\
 Onsala2           &  0.261 (0.009) & 0.260& -0.001\\
 G 12.89$+$0.49    &  0.428 (0.022) & 0.423& -0.005\\
 G 15.03$-$0.67    &  0.505 (0.033) & 0.495& -0.010\\
 G 192.16$-$3.84   &  0.660 (0.040) & 0.652& -0.008\\
 G 75.30$+$1.32    &  0.108 (0.005) & 0.108&  0.000\\
   W75 N           &  0.772 (0.042) & 0.763& -0.009\\
   DR 21           &  0.666 (0.035) & 0.659& -0.007\\
   DR 20           &  0.687 (0.038) & 0.678& -0.009\\
 IRAS 20290$+$4052 &  0.737 (0.062) & 0.715& -0.022\\
 AFGL 2591         &  0.300 (0.010) & 0.299& -0.001\\
\hline
 \end{tabular}
\end{center}
\end{table}
\begin{table*} []
 \begin{center}
  \centerline {Table 2 (the end)}
  \small
  \begin{tabular}{|l|r|r|r|}\hline
             Source    & $\varpi_{obs}$ & $\varpi_{corr}$&Correction\\\hline
 HW9 CepA              &  1.430 (0.070) & 1.417& -0.013\\
  Sgr B2N              &  0.128 (0.015) & 0.122& -0.006\\
  Sgr B2M              &  0.130 (0.012) & 0.126& -0.004\\
 IRAS 05137$+$3919     &  0.086 (0.027) & 0.093&  0.007\\
 G 27.36$-$0.16        &  0.125 (0.042) & 0.103& -0.022\\
 G 48.61$+$0.02        &  0.199 (0.007) & 0.198& -0.001\\
 IRAS 20126$+$4104     &  0.610 (0.020) & 0.608& -0.002\\
 G 9.62$+$0.19         &  0.194 (0.023) & 0.177& -0.017\\
 MSXDC G034.43$+$00.24 &  0.643 (0.049) & 0.626& -0.017\\
 G 23.66$-$0.13        &  0.313 (0.039) & 0.286& -0.027\\
 EC 95                 &  2.410 (0.020) & 2.409& -0.001\\ \hline
 \end{tabular}
\end{center}
\end{table*}

\end{document}